\documentclass[3p]{elsarticle}

\usepackage{amsmath,amsfonts,graphics}

\begin{document}

\begin{frontmatter}

\title{Semi-Poisson nearest-neighbor distance statistics for a 2D dissipative map} 

\author{Jamal Sakhr}
\address{Department of Physics and Astronomy, The University of Western Ontario, London, Ontario, Canada N6A 3K7}
\ead{jsakhr@uwo.ca}

\begin{abstract}
Typical pseudotrajectories of 2D ergodic maps are known to possess Wignerian nearest-neighbor distance distributions (NNDDs). In the case of 2D chaotic dissipative maps, bounded aperiodic pseudotrajectories typically evolve on planar strange attractors. In this microarticle, the hypothesis that such pseudotrajectories should possess NNDDs that are intermediate between the Poisson and Wigner distributions is put forward and a rare example for which the intermediate distribution can be clearly identified is presented. In particular, it is demonstrated numerically that typical  pseudotrajectories evolving in the strange attractor of the standard 2D Ikeda map possess semi-Poissionian NNDDs.  
\end{abstract}

\begin{keyword}
2D chaotic maps \sep 2D dissipative maps \sep chaotic trajectories \sep geometrical statistics \sep nearest-neighbor distances \sep semi-Poisson distribution    
\end{keyword}

\end{frontmatter}

\section{Introduction} 

The study of two-dimensional (2D) discrete maps is fundamental to the field of nonlinear dynamics (see, for example, Ref.~\cite{lieber}). The chaotic trajectories of such maps can have a very intricate and complex spatial structure, and a key theoretical issue in nonlinear dynamics is the characterization of this spatial complexity. While the mathematical tools of fractal geometry \cite{Falconer} can, in part, quantify this kind of complexity, there are many geometrical features of chaotic trajectories that cannot be understood or described using differential or fractal geometry. A different and surprisingly  untapped approach to addressing this issue of spatial complexity, which is largely unknown to dynamicists, is the application of ideas and techniques from spatial and geometrical statistics \cite{spatstats,Rips,StoyandStoy,StoyandStoy2}. For example, spatial analysts often describe the geometrical structure of a spatially complex object by referring to the nature of its so-called ``distance characteristics''. The distribution of interpoint nearest-neighbor distances is perhaps the most rudimentary example of a ``distance characteristic'' and it is often used as a basic geometrical-statistical descriptor for spatial point sets arising from both theoretical models and physical data \cite{spatstats,StoyandStoy2}. 

For a 2D chaotic map that is ergodic, aperiodic trajectories densely and uniformly cover the full phase space of the map (or positive-measure subsets thereof). In this case, a typical pseudotrajectory (i.e., numerical trajectory) of the map spatially mimics a 2D Poisson point process and its (mean-scaled) Euclidean nearest-neighbor distance distribution (NNDD) should be well modeled by the Wigner distribution \cite{mePLA}. Chaotic maps are of course not necessarily ergodic. In fact, most of the well-known chaotic maps \cite{JCS03} are not ergodic. Thus, in general, bounded aperiodic pseudotrajectories of 2D chaotic maps will not have Wignerian NNDDs. 

\section{Rationale and Hypothesis} 

In the important case of 2D chaotic \emph{dissipative} maps bounded aperiodic pseudotrajectories typically evolve on complicated subsets of the phase space having non-integer dimension (i.e., ``strange sets''). Even if such pseudotrajectories were to uniformly cover a strange set (a feature specific to ergodic strange attractors), they should not be expected to have Wignerian NNDDs since their spatial structure will no longer be well modeled by a homogeneous 2D Poisson point process. A more appropriate model would be some kind of inhomogeneous point process on a planar strange attractor. From a spatial statistical modeling point of view, developing such a model in a general way is a challenge since an apt model must properly account for pseudotrajectories (generally) non-uniform coverage of the attractor and the fact that any self-similarity possessed by the attractor is approximate (e.g., quasi, statistical, etc.) rather than exact. The sought-after model would therefore be akin to but not the same as the model developed in Ref.~\cite{Pdspaper}, which offers one approach to extending the multi-dimensional homogeneous Poisson point process to fractal sets. 
For a generic 2D chaotic dissipative map, bounded aperiodic pseudotrajectories typically evolve on a strange set with non-integer dimension $d_f\in(1,2)$. The hypothesis, based on the general ideas of Ref.~\cite{Pdspaper}, is that the (mean-scaled) NNDD of such pseudotrajectories should be intermediate between the Poisson and Wigner distributions. In the special case where the pseudotrajectories \emph{uniformly} cover an attractor that is \emph{exactly} self-similar, the model of Ref.~\cite{Pdspaper} is directly applicable and the pertinent distribution is the Brody distribution. In general, however, the intermediate distribution is not known, and hence the need for an appropriate point process model. In the following, a rare example for which the intermediate distribution can be clearly identified is presented. 

\section{Example} 

\begin{figure}
\scalebox{0.373}{\includegraphics*{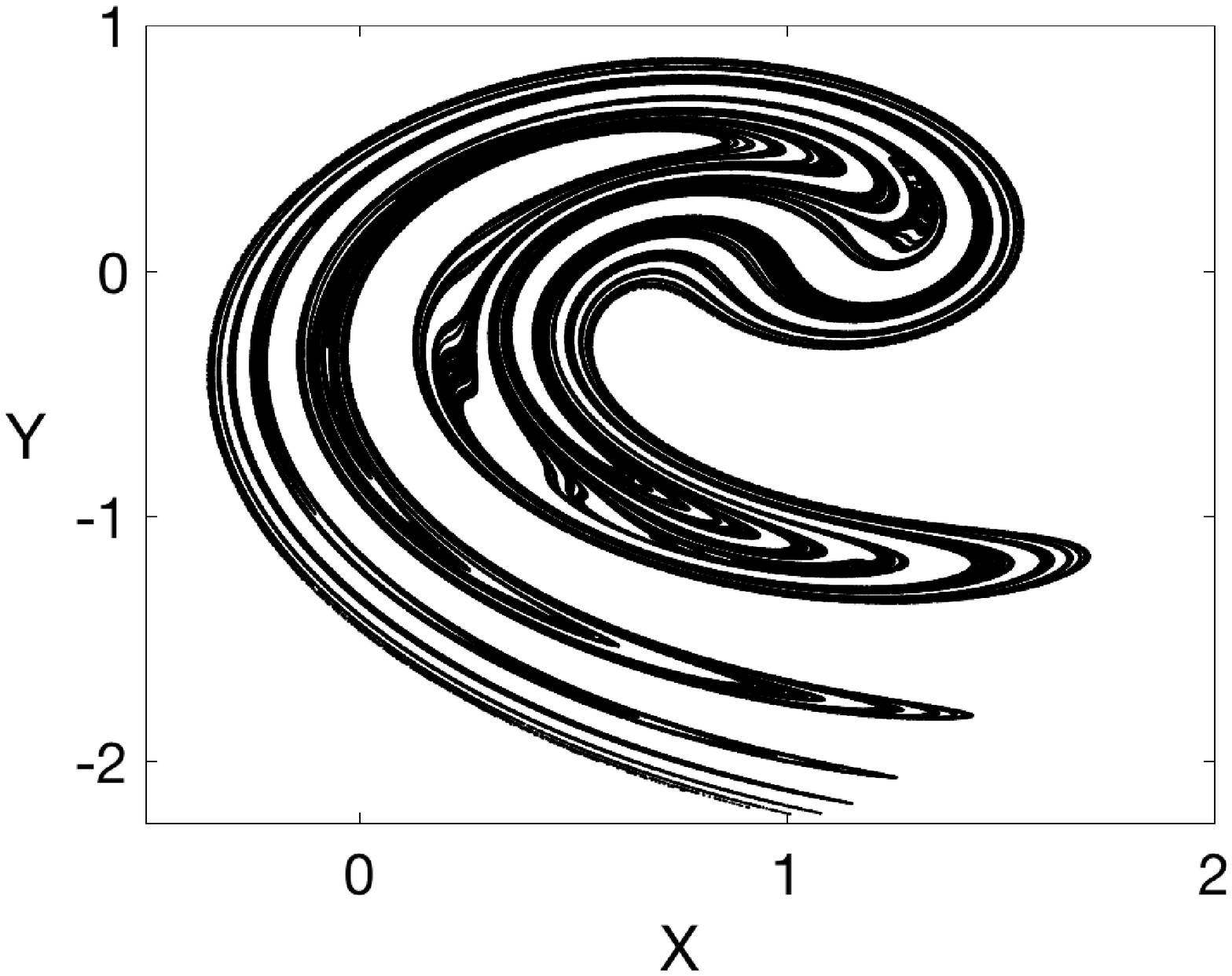}} 
\hspace*{0.5cm}
\scalebox{0.403}{\includegraphics*{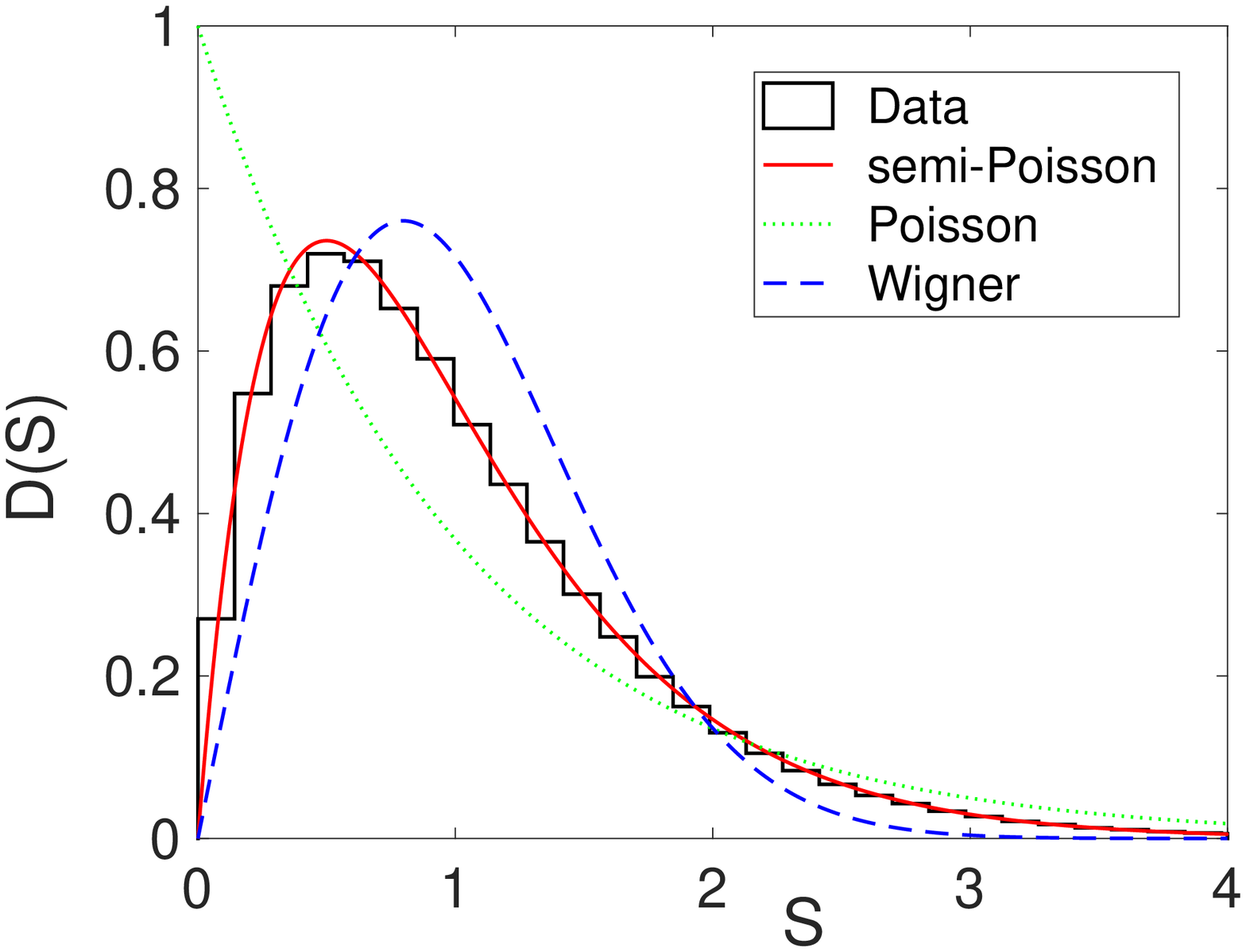}}
\caption{\label{ikedaresults} (Left) A typical pseudotrajectory in the strange attractor of the 2D Ikeda map (\ref{IKEDAMAP}) with standard parameter values A=1.0, B=0.9, C=0.4, and D=6.0. (Right) Density histogram of the interpoint nearest-neighbor distances for the pseudotrajectory shown to the left. The smooth red curve is the semi-Poisson distribution [Eq.~(\ref{semipoisson})].}
\end{figure}

Consider the 2D Ikeda map \cite{JCS03,Ikeda79}:
\begin{subequations}\label{IKEDAMAP}
\begin{equation}\label{IKEDAMAP1}
x_{n+1}=A+B\big(x_n\cos\phi_n-y_n\sin\phi_n\big),
\end{equation}
\begin{equation}\label{IKEDAMAP2}
y_{n+1}=B\big(x_n\sin\phi_n+y_n\cos\phi_n\big),
\end{equation}
where
\begin{equation}\label{IKEDAMAP3}
\phi_n=C-{D\over\big(1+x^2_n+y^2_n\big)},
\end{equation}
\end{subequations}
$n$ is a positive integer, and $\{A,B,C,D\}$ are tunable real-valued parameters. At the standard parameter values $\{A=1.0,~B=0.9,~C=0.4,~D=6.0\}$, the above map is dissipative and possesses a strange attractor having a fractal dimension of about 1.7 \cite{theiller90}. A typical pseudotrajectory in the attractor, of length $10^6$, is shown in the left panel of Fig.~\ref{ikedaresults}. This pseudotrajectory was launched from the initial point $(x_1,y_1)=(0,0)$; map (\ref{IKEDAMAP}) was iterated $1.1\times10^6$ times, and the first $10^5$ iterates discarded (to ensure convergence onto the attractor). The density histogram of the (mean-scaled) nearest-neighbor distances is shown in the right panel of Fig.~\ref{ikedaresults}. (The procedure for computing the mean-scaled nearest-neighbor distances is detailed in Ref.~\cite{mePLA}.) As conjectured, the latter lies in between the Poisson and Wigner distributions. In this particular case, the histogram data is remarkably well-modeled by the semi-Poisson distribution \cite{Bogo}:  
\begin{equation}\label{semipoisson}
D_{sP}(S)=4S\exp(-2S),
\end{equation}
where, in the present context, the variable $S\equiv z/\bar{z}$, where $z$ and $\bar{z}$ are the probability density variables for the nearest-neighbor distance and the sample mean nearest-neighbor distance, respectively. It is worth mentioning that the corresponding density histogram generated from the set of $10^5$ discarded iterates (which constitutes a distinct pseudotrajectory in its own right) is also well-modeled by the semi-Poisson distribution. 
Note that the (Wigner-like) linear behavior at small $S$ and the (Poisson-like) exponential behavior at large $S$ are clearly incompatible with and cannot be reproduced by the Brody distribution. 

\section{Conclusion} 

Chaotic pseudotrajectories of dissipative discrete maps generally have complex spatial structures. Characterizing and describing this spatial complexity is one of the main goals of nonlinear dynamics. The spatial statistical properties of such pseudotrajectories, which can serve as basic descriptors of their spatial structure, have yet to be formally studied and are completely unknown. This microarticle gives a rare example of a chaotic dissipative map whose pseudotrajectories possess at least one clearly identifiable spatial statistical property. More specifically, it was numerically demonstrated that typical pseudotrajectories evolving in the attractor of the standard 2D Ikeda map possess semi-Poissonian NNDDs. This interesting result supports the author's more general hypothesis that chaotic pseudotrajectories evolving in planar strange attractors possess (mean-scaled) NNDDs that are intermediate between the Poisson and Wigner distributions. Given the ubiquitous existence of strange attractors in dissipative nonlinear mapping systems and their generic spatial complexity, it would be very useful to develop an apt point process model that provides a comprehensive family of NNDDs that could characterize the nearest-neighbor distance statistics of pseudotrajectories evolving in any given planar strange attractor. For this hard task, the present hypothesis can serve as a helpful guide. 


\end{document}